\documentclass{article}

\usepackage{microtype}
\usepackage{graphicx}

\usepackage{hyperref}

\usepackage[accepted]{icml2026}

\usepackage{amsmath}
\usepackage{amssymb}

\usepackage[breakable,skins]{tcolorbox}
\usepackage{enumitem}
\newtcolorbox{promptbox}{
  enhanced,
  breakable,
  colback=gray!5,      
  colframe=gray!30,    
  arc=1mm,             
  boxrule=0.5pt,       
  left=10pt, right=10pt, top=10pt, bottom=10pt,
  fontupper=\small\ttfamily 
}
\usepackage{booktabs,longtable,array}

\newcommand{\pms}{\ensuremath{\pm}}

\icmltitlerunning{Proceedings of the ICML 2026 Workshop on Culture × AI, Seoul, Korea}

\begin{document}

\twocolumn[
  \icmltitle{What Gets Lost When Memory Becomes Media? Evaluating AI-Generated Oral History Visualization}



  \icmlsetsymbol{equal}{*}
  \icmlsetsymbol{second}{\dag}
  
  \begin{icmlauthorlist}
    \icmlauthor{Kwangsuk Park}{equal,aalab,aiffel}
    \icmlauthor{Jaehyun Koo}{equal,aalab,aiffel}
    \icmlauthor{Jiyeon Lee}{equal,aalab,aiffel}
    \icmlauthor{Anjung Tan}{second,aalab,promvalley}
    \icmlauthor{Hyoungchul Park}{second,aalab,aiffel}

  \end{icmlauthorlist}

  \icmlaffiliation{aalab}{AA LAB, MODULABS}
  \icmlaffiliation{aiffel}{Aiffel, MODULABS}
  \icmlaffiliation{promvalley}{New Business Development Team, PromValley}

  \icmlcorrespondingauthor{Kwangsuk Park}{kspark@MODULABS.co.kr}

  \icmlkeywords{CultureXAI, ICML}

  \vskip 0.3in
]



\printAffiliationsAndNotice{}  

\begin{abstract}
What gets lost when memory becomes media? Diaspora oral-history interviews require a double transformation; first-person recollection to third-person scene, present interview room to past time and place. When generative AI performs this transformation, no agreed criteria for success exist. We derive success conditions from oral-history theory, design 15 metrics around three failure modes, and compare a Multi-Agent Scene-decomposition pipeline (MAS) with a Single Summarization Pipeline (SSP) across 82 interviews from diaspora communities, spanning from oral interviews to 6-image sequences. Scene-planning and narrative preservation conflict  in the majority of cases, and the narrative-structure strength of the source testimony is the primary predictor of this conflict. We propose a failure-mode-based evaluation framework, an empirical analysis of conflict conditions, and a routing protocol for system selection based on narrative-structure strength.
\end{abstract}
 
\section{Introduction}

Diaspora oral-history interviews are recorded as present-day speakers recalling the past. A seventy-year-old sits before the camera, but the testimony points to a harbor in the 1950s or a classroom in the 1980s. Transforming such testimony into visual media means converting first-person recollection into third-person scenes, reconstructing period-specific settings, and producing a multi-panel image sequence. Visual media can convey the embodied, spatial, and temporal dimensions of these memories in ways that text summaries cannot, making image sequences a natural target format for this task.

Multiple values must be pursued simultaneously. Portelli~\cite{portelli1991trastulli} argues that the way a speaker arranges and transitions between events is itself historical evidence; Hirsch~\cite{hirsch2012postmemory} shows that diaspora memory is transmitted through material cues such as specific objects, recurring questions, and unfinished sentences. At the same time, the foundational ethics of oral history~\cite{abrams2010oral,warren2013restoring} require that the factual backbone not be distorted. These values can conflict: emphasizing scene-level texture may weaken macro narrative, while prioritizing factual accuracy may dissolve the sensory specificity of memory. This tension is well known in oral-history scholarship~\cite{portelli1991trastulli,abrams2010oral}, but how it manifests in media generation has not yet been systematically observed.

Observing this tension requires two approaches that emphasize different values. We implement a Single Summarization Pipeline (SSP) that compresses testimony into a single continuous narrative, and a Multi-Agent Scene-decomposition pipeline (MAS) that decomposes testimony into individual memory episodes rendered as scenes. We compare them on 82 interviews (each approximately one hour, with a key three-minute segment extracted) from diaspora communities. Three findings emerged. First, scene-planning and preservation conflict in the majority cases, indicating a structural property of the task. Second, the strength of the narrative-structure of the source testimony is the primary predictor of this conflict. Third, this pattern is invisible under aggregated scoring and visible only under value-separated evaluation.

Our primary contributions are as follows:
\begin{itemize} 
    \item an evaluation framework with 15 metrics derived from three oral-history failure modes, reported separately by failure mode 
    \item an empirical analysis showing that scene-preservation conflict occurs structurally and is predicted by testimony narrative-structure strength
    \item  a routing protocol for selecting SSP or MAS based on narrative-structure strength.
\end{itemize}


\section{Case Systems}
 
\begin{figure}[t]
\centering
\includegraphics[width=\columnwidth]{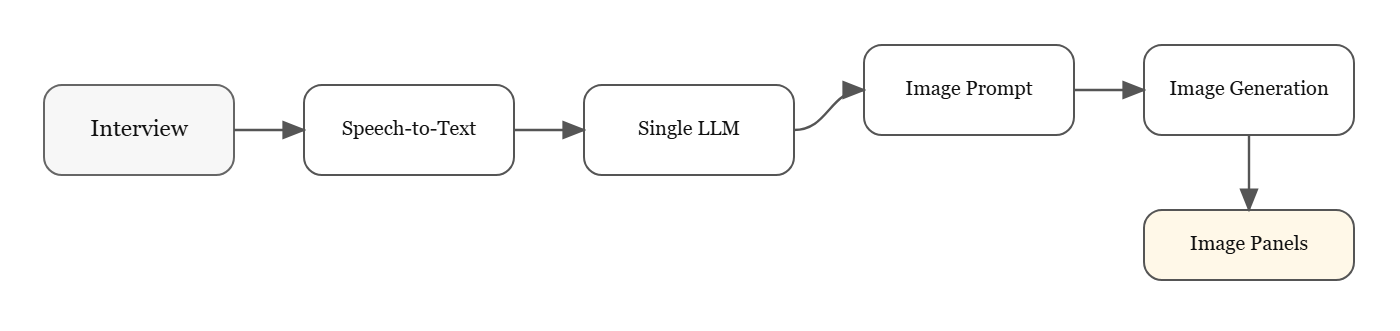}
\caption{SSP pipeline.}
\label{fig:pipeline_ssp}
\end{figure}

\begin{figure}[t]
\centering
\includegraphics[width=\columnwidth]{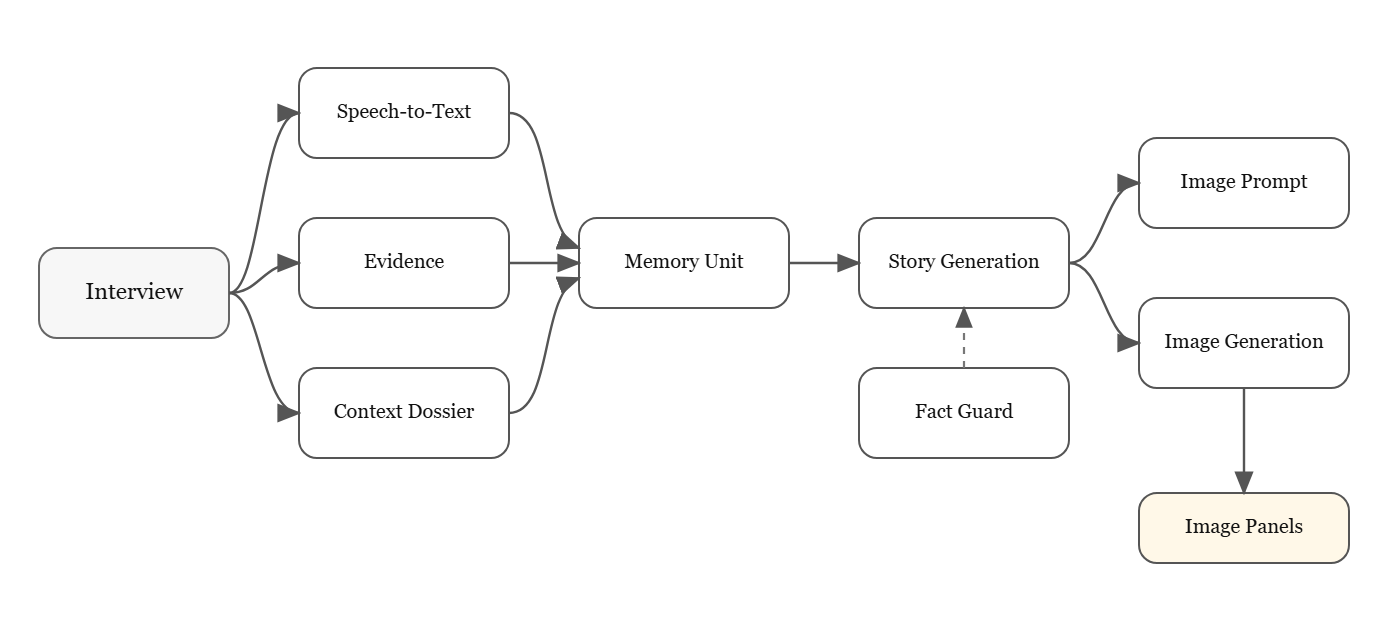}
\caption{MAS pipeline.}
\label{fig:pipeline_mas}
\end{figure}

Both systems take the same input, a recorded oral-history interview, and produce the same output, a six-panel image sequence. The shared path is: extract text from the interview, reconstruct the first-person account as a third-person period narrative, and sequentially generate six representative images. The difference lies in the structure of the intermediate representation (Figure~\ref{fig:pipeline_ssp},Figure~\ref{fig:pipeline_mas}).
 
\subsection{SSP: Single Summarization Pipeline}
 
SSP is a preservation-oriented design that compresses the entire testimony into a single continuous narrative. It extracts text from the interview, and a single LLM reconstructs it as a third-person summary narrative. Point-of-view conversion, period reconstruction, and scene selection are handled implicitly within one model call. Image prompts are then extracted from the narrative and used to generate the image sequence.
 
\subsection{MAS: Multi-Agent Scene-decomposition Pipeline}
 
Role-specialized agent architectures that distribute evaluation across modalities have improved consistency and alignment with structured criteria in domains such as code evaluation \cite{park2025agacci} and math problem generation \cite{lee2024vista}. MAS applies a similar modality-aware modular design to oral-history visualization, explicitly decomposing the double transformation, point-of-view conversion and period reconstruction. Text is extracted from the interview while, in parallel, an Evidence agent identifies emotional turning points and memory emphasis from the interview audio and a Context Dossier reconstructs period, regional, and cultural context. These three inputs are integrated and decomposed into 6 memory episodes in the Memory Unit. Story Generation then reconstructs each episode as a third-person scene narrative set in its historical period. A Fact Guard checks factual boundaries, and image prompts and images are generated per scene.
 
MAS separates into distinct stages what SSP handles implicitly in a single model call: point-of-view conversion, period reconstruction, scene selection, and factual checking. This makes each interpretive decision observable.
 
\subsection{Implementation}
 
Transcription uses Whisper~\cite{radford2023robust}. Story generation and all MAS text processing use GPT-5 mini. Both systems share Gemini 2.5 Flash Image~\cite{google2025gemini25flashimage} for image generation. Agent orchestration uses AutoGen~\cite{wu2024autogen}.
 
Because the two systems share the same image-generation model and conditions, differences in image output originate in the structure of the intermediate representation.
 
\section{Evaluation Framework}
 
SSP targets preservation of narrative flow; MAS targets surfacing of scene-level texture. Strengthening preservation can abstract away sensory detail; strengthening scene-level rendering can weaken narrative flow and factual context. The two values structurally conflict.
 
\subsection{Failure Modes}
 
Inverting the success conditions derived from oral-history theory yields three failure modes.
 
\textbf{Transition Dissolution (FM1)} $\leftrightarrow$ preservation of memory-transition structure. Portelli~\cite{portelli1991trastulli} shows that how a speaker arranges and transitions between events is itself evidence. A single summary can merge multiple memory episodes into one smooth paragraph, erasing transition points. In image-panel generation, these transitions become panel boundaries, so their loss affects the entire visualization.
 
\textbf{Genericization (FM2)} $\leftrightarrow$ surfacing of testimony-specific cues. Hirsch~\cite{hirsch2012postmemory} shows that memory is transmitted through material mediators, not abstract messages. IImage-generation models tend to converge on generic refugee visual conventions, such as isolated figures, borders, dark rooms~\cite{eccariuskelly2022digital,saltsman2021storytelling}. When testimony-specific cues are replaced by these conventions, the mediators of memory are lost.
 
\textbf{Preservation Damage (FM3)} $\leftrightarrow$ faithful reconstruction of historical context. The foundational ethics of oral history require respecting the speaker's memory while not distorting factual boundaries~\cite{abrams2010oral,warren2013restoring}. If scene-making damages macro narrative flow, destabilizes speaker identity and context, or distorts hard factual anchors, ethical boundaries are crossed. The double transformation, namely point-of-view conversion plus period reconstruction, heightens this risk.
 
\subsection{Metrics}
 
We designed 15 metrics to measure these three failure modes across text and image modalities as shwon at Table~\ref{tab:matrix} and Table~\ref{tab:defs}. Text metrics evaluate the quality of the intermediate representation for image generation; image metrics evaluate whether the final visual artifact preserves testimony specificity.
 
\begin{table}[t]
\centering
\caption{Metric matrix by failure mode (T = text, I = image).}
\label{tab:matrix}
\small
\begin{tabular}{lll}
\toprule
\textbf{FM} & \textbf{Mod.} & \textbf{Metric} \\
\midrule
FM1 & T & Panel Transition Clarity (PTC) \\
FM1 & T & Visual Scene Extractability (VSE) \\
FM1 & I & Narrative Progression (NP) \\
FM1 & I & Memory Scene Shift (MSS) \\
FM1 & I & Narrative Action (NA) \\
\midrule
FM2 & I & Non-Genericity (NG) \\
FM2 & I & Testimony-Specific Scene Detail (TSD) \\
FM2 & I & Visual Evidence Coverage (VEC) \\
\midrule
FM3 & T & Macro Arc Fidelity (MAF) \\
FM3 & T & Identity Context Lock (ICL) \\
FM3 & T & Testimony Grounding (TG) \\
FM3 & T & Hard Anchor Safety (HAS) \\
FM3 & I & Identity Continuity (IC) \\
FM3 & I & Period/Region Plausibility (PRP) \\
\midrule
Bnd & T & Allowable Fiction Control (AFC) \\
\bottomrule
\end{tabular}
\end{table}

\begin{table}[t]
\centering
\caption{Metric definitions.}
\label{tab:defs}
\small
\begin{tabular}{ll}
\toprule
\textbf{Abbr.} & \textbf{Measures} \\
\midrule
PTC & Preservation vs.\ dissolution of episode transitions \\
VSE & Extractability of image-ready scene cues \\
NP  & Panels form a memory flow, not isolated stills \\
MSS & Shift from interview to recalled past scene \\
NA  & Action and situational change vs.\ static portraits \\
\midrule
NG  & Testimony-specific features vs.\ refugee clich\'{e}s \\
TSD & Presence of specific objects, places, gestures \\
VEC & Breadth of visual evidence across panels \\
\midrule
MAF & Preservation of event flow and causal order \\
ICL & Stability of speaker identity and cultural context \\
TG  & Grounding in source testimony content and emphasis \\
HAS & Non-distortion of dates, places, names, relations \\
IC  & Cross-panel consistency of protagonist appearance \\
PRP & Period and regional accuracy of dress, architecture \\
\midrule
AFC & Sensory embellishment stays within testimony scope \\
\bottomrule
\end{tabular}
\end{table}
 
FM1 metrics track whether memory-transition points survive as panel boundaries in both text and image. FM2 metrics check whether the material memory mediators identified by Hirsch~\cite{hirsch2012postmemory} are replaced by generic imagery. FM3 metrics monitor whether narrative, identity, and factual anchors are damaged during point-of-view conversion and period reconstruction. AFC separately monitors whether sensory embellishment for image generation stays within the testimony's scope.
 
\subsection{Protocol}
 
We use grouped G-Eval~\cite{liu2023geval} with Gemini 2.5 Flash~\cite{comanici2025gemini} as judge. For text, the judge sees both conditions for the same interview and rates each metric on a 1--5 scale. For images, it compares same-index panels across conditions. Each unit is evaluated 10 times; we report mean $\pm$ SD and win rate.
 
We use 82 interviews from the RFMI Oral History Archive~\cite{rfmiArchive}, each approximately one hour, with a key three-minute segment extracted. From each interview, SSP and MAS each extract six key scene stories and generate a corresponding six-image sequence, yielding 164 text narratives and 984 images. At the image-generation stage, safety filters blocked images for some interviews; 66 interviews with complete six-image sequences were used for image evaluation. Across text and image, approximately 11{,}000 judge evaluations were conducted.
 
\section{Results}
 
\begin{figure}[t]
\centering
\includegraphics[width=\columnwidth]{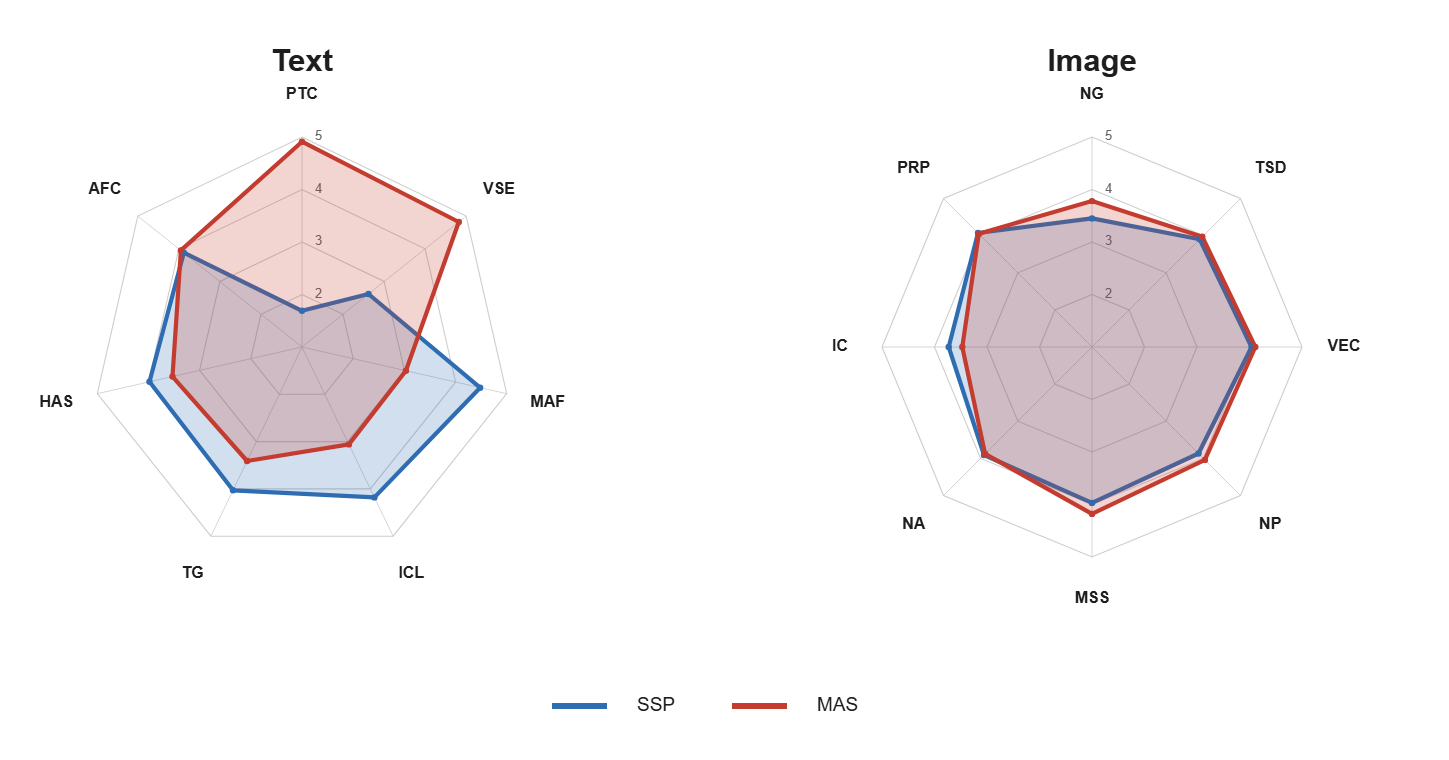}
\caption{Value profiles. Left: text (7 metrics). SSP (blue) and MAS (orange) extend in opposite directions. Right: image (8 metrics).}
\label{fig:radar}
\end{figure}

The two systems' metric averages are nearly identical, but their per-metric profiles are opposite (Figure~\ref{fig:radar}).
 
\subsection{Text: Scenes Sharpen, Narrative Weakens}
 
\begin{table}[t]
\centering
\caption{Text evaluation. 82 interviews, 1--5 scale, mean $\pm$ SD, G-Eval $\times$10.}
\label{tab:text}
\small
\begin{tabular}{llccc}
\toprule
\textbf{FM} & \textbf{Metric} & \textbf{SSP} & \textbf{MAS} & \textbf{Win} \\
\midrule
FM1 & PTC & 1.69\pms0.71 & \textbf{4.91\pms0.38} & MAS 97\% \\
FM1 & VSE & 2.62\pms0.75 & \textbf{4.82\pms0.53} & MAS 93\% \\
\midrule
FM3 & MAF & \textbf{4.48\pms0.85} & 3.03\pms1.21 & SSP 70\% \\
FM3 & ICL & \textbf{4.18\pms1.05} & 3.06\pms0.97 & SSP 74\% \\
FM3 & TG  & \textbf{4.03\pms1.27} & 3.41\pms1.22 & SSP 62\% \\
FM3 & HAS & \textbf{3.98\pms1.32} & 3.53\pms1.33 & SSP 53\% \\
\midrule
Bnd & AFC & 3.88\pms1.34 & 3.95\pms1.22 & tie \\
\bottomrule
\end{tabular}
\end{table}
 
That MAS dominates FM1 metrics is expected by design. What was not predicted is the magnitude of FM3 loss: MAF and ICL both show SSP leading by over one point on a five-point scale, with win rates above 70\%. MAS scene-decomposition incurs substantial costs to macro narrative flow and identity-context stability.
 
AFC is nearly tied, with a marginal difference of $+$0.07. Contrary to the expectation that MAS would fabricate more, the two systems produce comparable amounts of embellishment in different directions: SSP smooths the overall narrative; MAS fills sensory detail in individual scenes. This directional difference is the mechanism that simultaneously produces FM1 gains and FM3 losses.
 
FM3 losses are not uniform across cases. SSP's MAF win rate of 70.1\% means that in roughly 30\% of cases MAS did not incur preservation loss. This case-level variation is analyzed in \S\ref{sec:conflict}.
 
\subsection{Image: Clich\'{e}s Decrease, Identity Wavers}
 
\begin{table}[t]
\centering
\caption{Image evaluation. 66 interviews (6-image sequences), 1--5 scale, mean $\pm$ SD, G-Eval $\times$10.}
\label{tab:image}
\small
\begin{tabular}{llccc}
\toprule
\textbf{FM} & \textbf{Metric} & \textbf{SSP} & \textbf{MAS} & \textbf{Win} \\
\midrule
FM2 & NG  & 3.45\pms0.75 & \textbf{3.78\pms0.77} & MAS 47\% \\
FM2 & TSD & 3.90\pms0.91 & \textbf{3.97\pms0.92} & MAS 36\% \\
FM2 & VEC & 4.04\pms0.87 & \textbf{4.11\pms0.87} & MAS 33\% \\
\midrule
FM1 & NP  & 3.87\pms0.88 & \textbf{4.04\pms0.81} & MAS 30\% \\
FM1 & MSS & 3.97\pms1.07 & \textbf{4.18\pms0.85} & MAS 17\% \\
FM1 & NA  & 3.91\pms0.70 & 3.88\pms0.69 & tie \\
\midrule
FM3 & IC  & \textbf{3.73\pms0.77} & 3.47\pms0.96 & SSP 87\% \\
FM3 & PRP & 4.07\pms0.66 & 4.05\pms0.59 & tie \\
\bottomrule
\end{tabular}
\end{table}
 
The dramatic FM1 gains in text are substantially attenuated in images. Text PTC and VSE advantages exceeded three points; image NP and MSS differences are around 0.2 points. The structural difference in the intermediate representation is largely absorbed by the image-generation model. Within this, MSS shows MAS better facilitating the shift from present interview to recalled past, serving as a meaningful signal for the double-transformation task. NA is essentially tied, suggesting that action depiction is governed by the image model itself rather than the intermediate representation.
 
FM2 metrics show modest MAS advantages, but no win rate reaches 50\%. MAS does not consistently produce less generic images in every case; it reduces generic convergence on average. The larger improvement in NG compared to TSD and VEC suggests MAS is more effective at avoiding clich\'{e}s than at filling their place with testimony-specific detail.
 
IC and PRP, both FM3, diverge: IC shows SSP clearly more stable (win rate 87\%), while PRP is tied. Period-appropriate settings (architecture, dress) persist across scene changes because they are specified in prompts, but character faces and physiques are regenerated per scene, making identity consistency harder to maintain. Period plausibility and action depiction appear governed by the image model's baseline capabilities rather than by intermediate-representation structure.
 
Notably, MAS shows lower standard deviations than SSP in MSS, PRP, and AFC, suggesting that explicit decomposition reduces case-to-case variance even where it does not shift the mean.
 
\subsection{Gains and Costs Propagate from Text to Image}
 
The text-level FM1-vs-FM3 trade-off repeats in images as FM2-vs-FM3. Clearer scene-transition structure in text leads the image model to incorporate more testimony-specific detail per scene, reducing generic convergence. At the same time, greater scene diversity propagates identity instability. This propagation is heavily attenuated compared to text-level differences, indicating that the image-generation model is a bottleneck that does not fully transmit intermediate-representation improvements.
 
\subsection{The Conflict Is Structural}
\label{sec:conflict}
 
The text radar profile (Figure~\ref{fig:radar}, left) shows clear shape divergence between SSP and MAS, while the image profile largely overlaps. We therefore focus case-level pattern analysis on the text metrics with the starkest contrast. For each of 82 cases, we compute deltas on the scene-planning axis (PTC, VSE) and the preservation axis (MAF, ICL, TG, HAS) and classify the pattern (Table~\ref{tab:pattern}).
 
\begin{table}[t]
\centering
\caption{Conflict-pattern distribution (82 text evaluations).}
\label{tab:pattern}
\small
\begin{tabular}{lrr}
\toprule
\textbf{Pattern} & \textbf{n} & \textbf{\%} \\
\midrule
Trade-off (scene$\uparrow$ preservation$\downarrow$) & 59 & 68.6 \\
MAS win (scene$\uparrow$ preservation$\uparrow$) & 24 & 27.9 \\
\quad of which all-metric win & 13 & 15.1 \\
SSP win (scene$\downarrow$ preservation$\uparrow$) & 2 & 2.3 \\
\quad of which all-metric win & 1 & 1.2 \\
\bottomrule
\end{tabular}
\end{table}
 
The trade-off appears in more than two out of three cases. This is not a defect of either system but a structural property of the task. In roughly one in four cases, MAS improved both scene-planning and preservation, and more than half of those (13 cases) showed MAS winning on every metric. When the trade-off resolves, it tends to resolve completely, suggesting that preservation loss is driven by a condition-specific mechanism rather than a systemic flaw.
 
\subsection{Conflict Conditions: Narrative-Structure Strength}
 
\begin{figure}[t]
\centering
\includegraphics[width=\columnwidth]{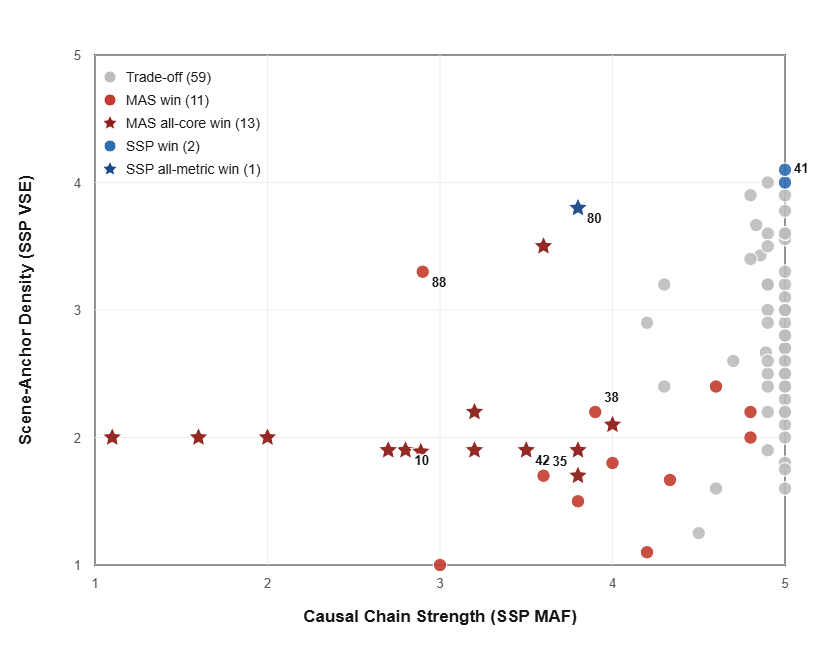}
\caption{Routing scatter. x: narrative-structure strength (SSP MAF). y: scene-anchor density (SSP VSE). Grey: trade-off. Red/star: MAS win. Blue: SSP win.}
\label{fig:scatter}
\end{figure}

To identify what separates trade-off cases from resolution cases, we define two proxies from existing metrics:
\begin{itemize}
\item \textbf{Narrative-structure strength} = case-level SSP MAF mean
\item \textbf{Scene-anchor density} = case-level SSP VSE mean
\end{itemize}
Because SSP compresses testimony as-is into a single narrative, its scores reflect the structural properties of the source testimony rather than the system's own capabilities. Figure~\ref{fig:scatter} plots these proxies for all 82 cases, colored by pattern.
 
The dominant pattern follows the x-axis. Trade-off cases concentrate on the right (high SSP MAF); MAS-win cases concentrate on the left (low SSP MAF). When the source testimony has strong narrative structure, SSP already preserves it well, and MAS decomposition damages what is already preserved as the more structure there is to decompose, the more there is to damage. When narrative structure is weak, SSP also fails to capture it, and MAS decomposition supplies missing structure, improving both scene-planning and preservation simultaneously.
 
Scene-anchor density (y-axis) does not separate patterns on its own. Even with rich scene cues, strong narrative structure causes trade-off because decomposition costs outweigh scene-planning gains.
 
\begin{figure}[t]
\centering
\includegraphics[width=\columnwidth]{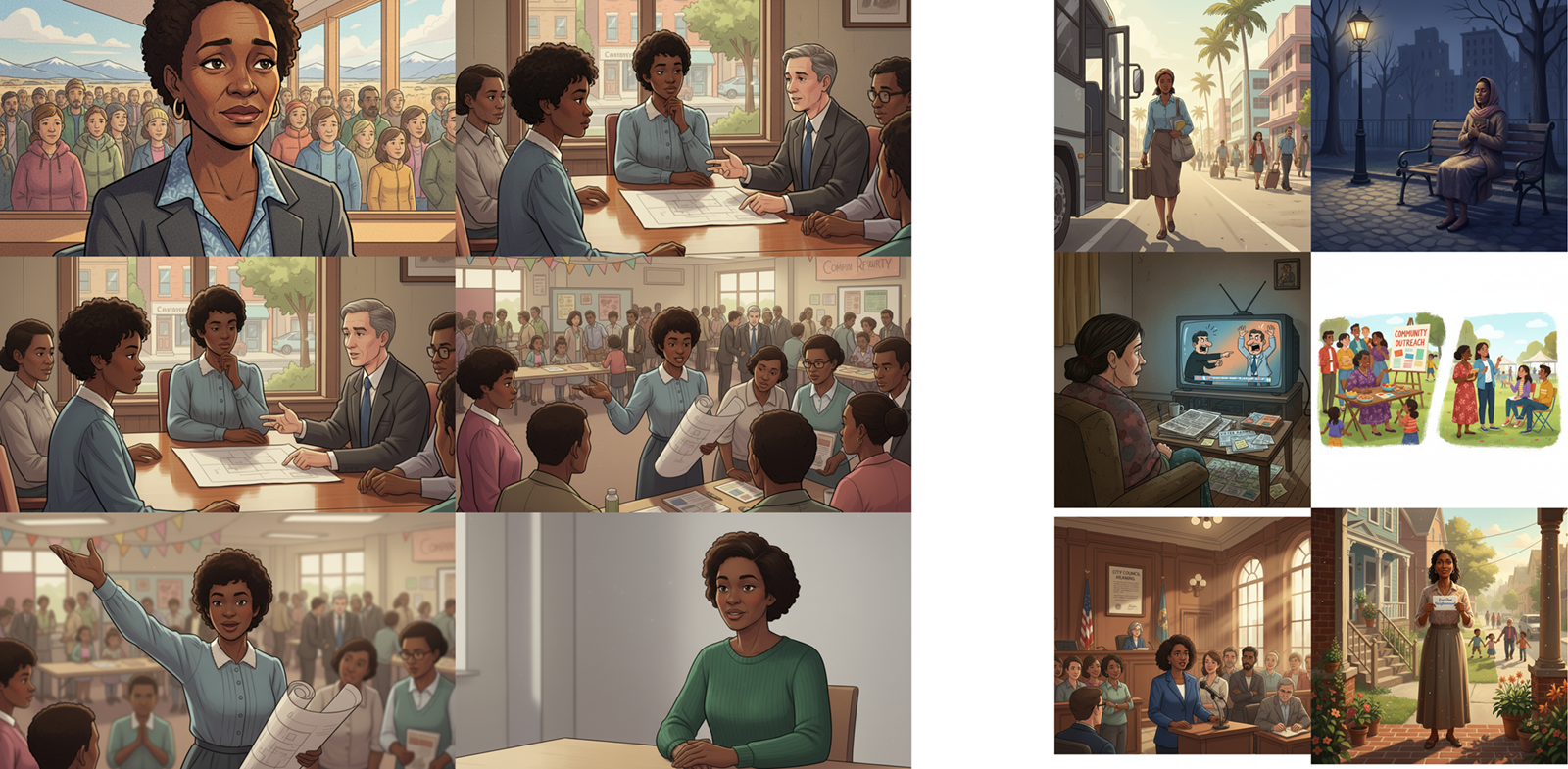}
\caption{Case 38 image comparison. Left: MAS. Right: SSP.}
\label{fig:case38}
\end{figure}

Qualitative cases confirm this pattern. Case~80 (strong causal chain: airport corruption $\to$ taxi warning $\to$ mother's flight $\to$ DEA convoy $\to$ karaoke-bar release) sits on the right of the scatter; SSP preserves the chain while MAS fragments it. Case~38 sits at moderate MAF yet shows MAS all-metric win (Figure~\ref{fig:case38}), indicating that narrative-structure strength is the primary but not sole predictor; qualitative properties of the testimony (types of sensory cues, speaker's recall style) act as additional variables. Cases~42, 35, 10, and 88 show similar patterns; detailed analysis is in Appendix~D.
 
\begin{table}[t]
\centering
\small
\begin{tabular}{p{\linewidth}p{\linewidth}}
\toprule
Compressed story \\
\midrule
Jorge Quintas recounts his family's perilous journey through Mexico, facing extortion and cartel kidnapping threats. Saved by his mother’s resourcefulness and DEA intervention, their story underscores the constant danger and corruption migrants navigate while seeking safety in the United States. \\
\bottomrule
\caption{Compressed story of case 38}
\end{tabular}
\end{table}

\begin{figure}[t]
\centering
\includegraphics[width=\columnwidth]{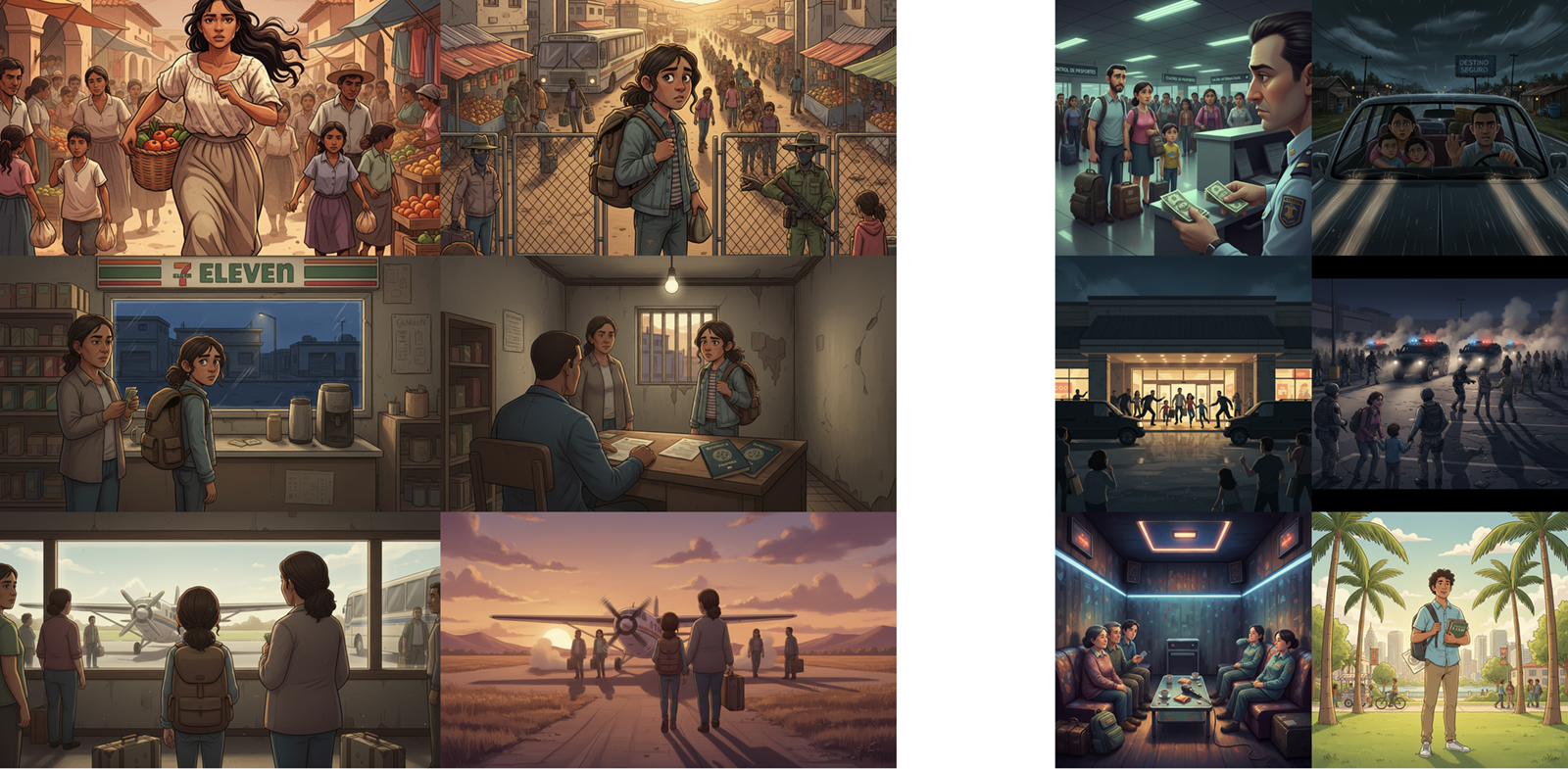}
\caption{Case 80 image comparison. Left: MAS. Right: SSP.}
\label{fig:case80}
\end{figure}

\begin{table}[t]
\centering
\small
\begin{tabular}{p{\linewidth}p{\linewidth}}
\toprule
Compressed story \\
\midrule
Darara Gubo's journey from an Ethiopian immigrant to a U.S. citizen has fueled her resolve to seek local office. Driven by her faith and past experiences with poverty, she transitioned from a cautious observer to a civic leader. Encouraged by her community, she now aims to serve on the city council, prioritizing grassroots neighborly service over distant political power. \\
\bottomrule
\caption{Compressed story of case 80}
\end{tabular}
\end{table}

\section{Routing Protocol}
 
\begin{figure}[t]
\centering
\includegraphics[width=\columnwidth]{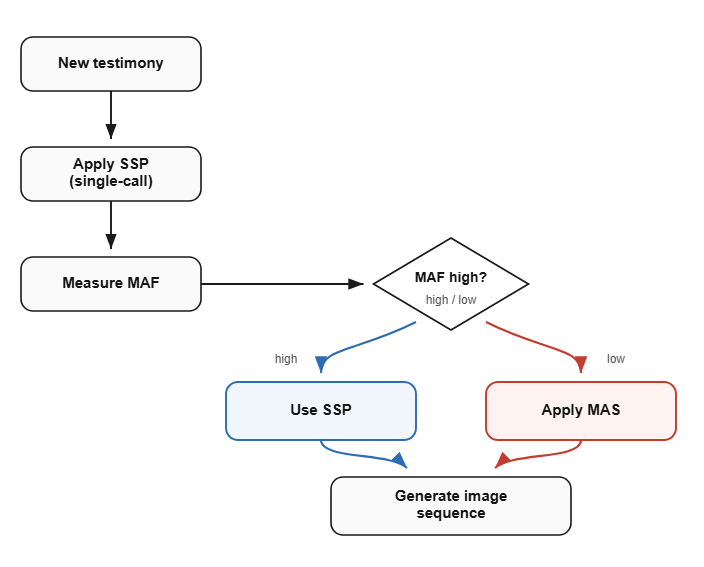}
\caption{Routing protocol. New testimony $\to$ SSP $\to$ MAF measurement $\to$ high: use SSP / low: apply MAS.}
\label{fig:routing}
\end{figure}

From the analysis in Section~4, narrative-structure strength (SSP MAF) is the primary predictor of whether scene-decomposition will produce a trade-off or a net gain. We propose a system-selection procedure based on this finding (Figure~\ref{fig:routing}).
 
Given a new testimony, first apply SSP and measure the MAF score of the resulting narrative. This score reflects the narrative-structure strength of the source testimony. If MAF is high, SSP already preserves the narrative structure well; use the SSP result, as MAS is likely to damage it. If MAF is low, SSP has not captured the structure adequately; apply MAS, whose explicit decomposition is likely to supply missing structure. In this dataset, the boundary is observed near MAF~3.0, though generalization of this threshold requires further validation.
 
\section{Discussion}
 
Our findings are clear. The conflict between scene-planning and preservation is not a system defect but a property inherent to oral-history visualization, and its occurrence is predicted by the narrative-structure strength of the source testimony. Just as oral history tradition located the value of oral history in the tension between factual accuracy and the subjective form of memory, this study shows that the same tension manifests in measurable form in the media-generation process. The metrics, conflict analysis, and routing protocol proposed here constitute a first structure for managing this tension.
 
Validation and refinement of this structure remain as areas for further study. Future work will include human evaluation by oral-history researchers and community members, optimization of the routing threshold, and addressing the conflict between safety filters and cultural record, as evidenced by the 16 of 82 interviews blocked during image generation.
 
Success in oral-history visualization is not a single score. It is a structure for measuring which values are strengthened and weakened, diagnosing where the conflict originates, and selecting an appropriate approach accordingly.

\section*{Impact Statement}

This research proposes 15 evaluation metrics to address narrative distortion and ethical issues in AI-generated oral history visualization. By establishing a systematic framework to minimize identity loss when converting diaspora memories into images, it contributes to the responsible preservation and visualization of historical records.


\bibliography{example_paper}
\bibliographystyle{icml2026}

\newpage
\appendix
\onecolumn
\section*{Appendix}
\section{Prompt Appendix}
\label{app:prompts}

This appendix reports the prompt templates used in the SSP/MAS comparison. We use the paper terminology throughout: \texttt{SSP} denotes the Single Summarization Pipeline and \texttt{MAS} denotes the Multi-Agent Scene-decomposition pipeline. The MAS prompt order follows Figure~1: Speech-to-Text, Evidence, Context Dossier, Memory Unit, Story Generation, Fact Guard, Image Prompt, and Image Generation. Internal artifact labels from earlier runs are omitted for readability. Bracketed fields such as \texttt{[transcript]} and \texttt{[metadata]} denote run-time variables populated from the interview record, transcript, MAS artifact, or image manifest. API keys and file-system paths are omitted.

\subsection{SSP Generation Prompts}
\label{app:ssp-prompts}

\subsubsection{SSP Text Generator}
The SSP text condition receives transcript text and metadata, then produces a length-matched third-person summary narrative. The target length is counted in code from the corresponding MAS story so that the text comparison does not reward one condition simply for being longer.

\begin{promptbox}
\obeylines\obeyspaces
You are the SSP Text Generator for a diaspora oral-history visualization study.
Produce a concise third-person summary narrative from transcript text and metadata only.
Do not create a scene-decomposition plan. Do not add camera language. Do not add image prompts.
Preserve chronology, speaker agency, core relationships, places, and emotional arc.
Allowed creative wording is limited to connective prose; do not invent events, countries, eras, family roles, or identity.
Target length: between [min\_chars] and [target\_chars] Python characters.
Output JSON only.

User payload:
\{
  "metadata": [metadata],
  "transcript": [transcript, compacted to 22000 characters],
  "target\_chars\_hard\_counted\_by\_code": [target\_chars],
  "output\_schema": \{
    "title": "short title",
    "summary\_en": "third-person testimony summary narrative",
    "summary\_ko": "Korean one-paragraph summary",
    "notes": ["short note"]
  \}
\}
\end{promptbox}

\subsubsection{Shared Image-Sequence Prompt Generator}
This prompt converts either the SSP narrative or the MAS story into six sequential image prompts without a specialized director stage. It is shared across the two image conditions so that the image comparison tests the source representation rather than a different image-prompting strategy.

\begin{promptbox}
\obeylines\obeyspaces
You are the Shared Image-Sequence Prompt Generator.
Convert the provided source text into [target\_panel\_count] sequential image prompts.
Each prompt should describe one distinct image in the sequence.
Do not create a specialized director plan.
Do not create identity-lock, fact-guard, or period-contract fields beyond what is explicitly present in the source text.
Keep prompts dignified, testimony-grounded, and visually concrete.
Output JSON only.

User payload:
\{
  "condition": "[SSP or MAS]",
  "metadata": [metadata],
  "source\_text": [SSP narrative or MAS story, compacted to 12000 characters],
  "target\_panel\_count": 6,
  "output\_schema": \{
    "image\_sequence": [
      \{
        "panel\_id": "[1..target\_panel\_count]",
        "summary": "what this image should show",
        "image\_prompt": "image generation prompt"
      \}
    ]
  \}
\}
\end{promptbox}

\subsubsection{Shared Image Prompt Wrapper}
Each sequential prompt is normalized by code before image generation. In the reported comparison, both SSP and MAS produce six image prompts per interview whenever image generation succeeds.

\begin{promptbox}
\obeylines\obeyspaces
Create one 16:9 cinematic non-photoreal 2D/2.5D illustrated still image.
No text, no captions, no UI, no split panels, no photoreal people, no real-person likeness.
Use a dignified diaspora oral-history memory style, not generic stock refugee imagery.
Keep the scene grounded in the provided source story; do not invent major events or identity changes.
[image\_prompt from the shared image-sequence prompt generator]
\end{promptbox}

\subsection{MAS Agent Prompts}
\label{app:mas-prompts}

\subsubsection{Speech-to-Text Agent}
The Speech-to-Text stage uses the transcription API directly and does not use a free-form natural-language prompt.

\begin{promptbox}
\obeylines\obeyspaces
Model: openai:gpt-4o-mini-transcribe
API task: audio transcription
Parameters:
- file: [audio chunk]
- language: [transcript\_language]
- response\_format: json
Output:
- chunk-level transcript text
- chunk start/end timestamps
\end{promptbox}

\subsubsection{Evidence Agent}
The Evidence Agent extracts testimony-grounded anchors before any scene reconstruction happens. This corresponds to the \textit{Evidence} box in Figure~1 and replaces earlier internal variants that separated factual, text-emotion, and video-emotion extraction.

\begin{promptbox}
\obeylines\obeyspaces
You are the Evidence Agent for a diaspora oral-history visualization system.
Use only the transcript and metadata.
Extract evidence that downstream agents must preserve or may use for scene construction.
Do not invent missing facts. If a detail is uncertain, mark it as uncertain.
Output JSON only.

User payload:
\{
  "metadata": [metadata],
  "transcript": [transcript],
  "output\_schema": \{
    "hard\_anchors": ["dates, places, people, family roles, routes, institutions"],
    "identity\_anchors": ["speaker identity and relational facts that must not drift"],
    "visual\_anchors": ["objects, rooms, streets, gestures, documents, clothing, vehicles"],
    "memory\_turning\_points": ["moments where the testimony shifts time, place, subject, or affect"],
    "emotion\_memory\_profile": [
      \{
        "evidence\_excerpt": "short transcript evidence",
        "emotion": "one label from [EMOTION\_LABELS]",
        "intensity\_0\_to\_1": 0.0,
        "memory\_weight": "low | medium | high"
      \}
    ],
    "must\_not\_claim": ["unsupported or forbidden inferences"],
    "uncertainties": ["details that require caveating"]
  \}
\}
\end{promptbox}

\subsubsection{Context Dossier Agent}
The Context Dossier Agent restores historical, geographic, and cultural context needed for downstream scene construction. It does not add new plot events.

\begin{promptbox}
\obeylines\obeyspaces
You are the Context Dossier Agent for a diaspora oral-history visualization system.
Use only the transcript, metadata, and extracted evidence.
Create a compact context dossier that downstream agents can use to avoid era, region, identity, and setting drift.
Do not invent missing facts. If a detail is uncertain, mark it as uncertain.
Output JSON only.

User payload:
\{
  "metadata": [metadata],
  "transcript": [transcript],
  "evidence": [evidence],
  "output\_schema": \{
    "origin\_context": "country, region, era, and displacement context",
    "resettlement\_context": "settlement location and period if available",
    "identity\_context": "speaker identity facts that must not drift",
    "period\_visual\_anchors": ["clothing, architecture, objects, social setting"],
    "forbidden\_inferences": ["facts or identity claims not supported by testimony"],
    "uncertainties": ["details that require caveating"]
  \}
\}
\end{promptbox}

\subsubsection{Memory Unit Agent}
The Memory Unit Agent partitions the transcript into memory episodes. This is the key MAS operation that makes transition points observable.

\begin{promptbox}
\obeylines\obeyspaces
You are the Memory Unit Agent.
Split the testimony into 3-8 memory episodes according to natural turning points.
Each episode should have a clear subject, setting, action, emotional weight, and evidence excerpt.
Do not force equal lengths. Do not split a tightly causal chain if the split would damage the macro arc.
Output JSON only.

User payload:
\{
  "transcript": [transcript],
  "evidence": [evidence],
  "context\_dossier": [context dossier],
  "output\_schema": \{
    "memory\_units": [
      \{
        "unit\_id": "[1..N]",
        "time\_range": "MM:SS-MM:SS if available",
        "episode\_summary": "what happens in this memory unit",
        "transition\_from\_previous": "why this is a new unit",
        "visual\_anchors": ["objects, places, gestures, people"],
        "macro\_arc\_role": "setup | displacement | rupture | adaptation | reflection | closure",
        "evidence\_excerpt": "verbatim or compact transcript evidence"
      \}
    ],
    "causal\_chain\_risk": "low | medium | high",
    "notes": ["short notes"]
  \}
\}
\end{promptbox}

\subsubsection{Story Generation Agent}
The Story Generation Agent converts the memory units into a third-person, scene-oriented story for image-sequence generation.

\begin{promptbox}
\obeylines\obeyspaces
You are the Story Generation Agent for diaspora oral-history visualization.
Transform the transcript-grounded memory units into a dignified third-person story in [story\_language].
Use one concise paragraph per memory episode.
Preserve factual spine, chronology, speaker agency, and emotional resonance.
Make scene transitions explicit so that each episode can become one image in a sequence.
When the testimony recalls a past event or third party, shift the scene into that remembered era/place.
Separate verified testimony from allowable visual reconstruction using must\_preserve and may\_imagine.
Avoid victim stereotypes, poverty spectacle, generic refugee imagery, and unsupported facts.
Output JSON only.
\end{promptbox}

\begin{promptbox}
\obeylines\obeyspaces
Story Generation Agent user payload:
\{
  "context\_dossier": [context dossier],
  "evidence": [evidence],
  "memory\_units": [memory units],
  "target\_story\_chars": [story\_chars],
  "output\_schema": \{
    "title": "short title",
    "story\_en": "one concise paragraph per story unit",
    "story\_metric\_text": "same-language factual summary for evaluation",
    "scene\_beats": [
      \{
        "beat\_id": "[1..N]",
        "summary": "planned scene summary",
        "evidence\_excerpt": "transcript-grounded evidence",
        "visual\_mode": "present\_interview | remembered\_reenactment | historical\_context | symbolic\_transition",
        "memory\_subject": "who the image follows",
        "identity\_lock": "short identity continuity contract",
        "scene\_era": "era or decade if known",
        "scene\_location": "location or region if known",
        "setting\_anchor": "concrete environment",
        "period\_visual\_contract": "era/location/clothing/architecture/object constraints",
        "must\_preserve": ["major testimony anchors that cannot change"],
        "may\_imagine": ["minor visual details allowed"],
        "creative\_license\_boundary": "what cannot be invented or changed",
        "visual\_action": "what the subject does",
        "environment\_change": "what changes around them"
      \}
    ]
  \}
\}
\end{promptbox}

\subsubsection{Fact Guard Agent}
The Fact Guard Agent checks the scene-oriented story before image prompting.

\begin{promptbox}
\obeylines\obeyspaces
You are the Fact Guard Agent.
Compare the reconstructed story and scene beats against the transcript and dossier.
Flag unsupported claims, identity drift, era drift, invented family roles, changed countries, or altered agency.
Do not penalize minor visual reconstruction if it remains within the testimony's large-scale factual boundary.
Output JSON only.

User payload:
\{
  "transcript": [transcript],
  "context\_dossier": [context dossier],
  "evidence": [evidence],
  "story": [MAS story],
  "scene\_beats": [scene beats],
  "output\_schema": \{
    "approved": true,
    "unsupported\_claims": ["empty if none"],
    "hard\_anchor\_risks": ["dates, places, names, relationships"],
    "identity\_context\_risks": ["identity, era, region, social context drift"],
    "revision\_notes": ["required changes if approved is false"]
  \}
\}
\end{promptbox}

\subsubsection{Image Prompt Agent}
The Image Prompt Agent converts each approved scene beat into one image prompt. It corresponds to the \textit{Image Prompt} box in Figure~1.

\begin{promptbox}
\obeylines\obeyspaces
You are the Image Prompt Agent.
Convert the approved MAS scene beat into one concise image-generation prompt.
The prompt must preserve the identity lock, period visual contract, evidence excerpt, visual action, and creative license boundary.
Output JSON only.

User payload:
\{
  "scene\_beat": [scene beat JSON],
  "fact\_guard": [fact guard result],
  "output\_schema": \{
    "panel\_id": "[1..target\_panel\_count]",
    "image\_summary": "one-sentence image summary",
    "image\_prompt": "final image prompt",
    "negative\_constraints": [
      "no text, captions, UI, labels, title cards",
      "no photoreal people, real-person likenesses, or live-action look",
      "no generic refugee stereotype, poverty spectacle, or sensationalized violence",
      "no unsupported identity, country, era, or family-role changes"
    ]
  \}
\}
\end{promptbox}

\subsubsection{Image Generation Agent}
The Image Generation Agent calls the image model with the Image Prompt Agent output. It does not use an additional reasoning prompt; the wrapper below is the final model-facing instruction.

\begin{promptbox}
\obeylines\obeyspaces
Create one 16:9 cinematic non-photoreal 2D/2.5D illustrated still image for a diaspora oral-history memory sequence.
Image prompt: [image prompt]
Scene beat: [scene beat JSON]
Identity lock: [identity\_lock]
Period visual contract: [period\_visual\_contract]
Evidence excerpt: [evidence\_excerpt]
Constraints:
- one full-frame image only, no split panels or comic grid
- no text, captions, UI, labels, or title-card design inside the image
- not a blank background, not an infographic, not stock documentary B-roll
- high-quality illustrated characters only; not photoreal people, not real-person likenesses, not live-action
- dignified depiction, no refugee stereotype, no poverty spectacle, no sensationalized violence
- follow the declared memory subject, scene era, scene location, and visual action
- never change the apparent ancestry, broad skin tone range, hair logic, or facial lineage of the same active character
- show a concrete action or environment change rather than a posed portrait
- include era/location-specific background activity and plausible period props
- preserve must\_preserve anchors and stay within the creative\_license\_boundary
\end{promptbox}

\subsection{Evaluation Prompts, Not Pipeline Agents}
\label{app:evaluation-prompts}

The following prompts are not part of the SSP or MAS generation pipelines. They are the grouped G-Eval prompts used after generation to score the reported text and image metrics.

\subsubsection{Text Grouped G-Eval Judge}
This grouped text evaluation prompt was used for the reported text metrics. The judge receives SSP and MAS outputs from the same interview in one call.

\begin{promptbox}
\obeylines\obeyspaces
You are a grouped G-Eval judge for testimony-grounded text used as an intermediate representation for visual memory reconstruction.
Evaluate SSP and MAS in the same call to reduce scale drift.
Do not reward generic fluency, smooth summarization, or polished prose by itself.

Procedure:
1. Identify the testimony's macro arc, hard anchors, protagonist identity, era/region context, and scene-level turning points.
2. Score each condition independently on every metric using the same standard.
3. Prefer text that is easier to convert into a faithful multi-image visual sequence only for scene-planning metrics.
4. For preservation metrics, prefer the condition that best preserves chronology, identity context, and hard factual anchors.
5. Creative connective narration is allowed only if it does not distort the main event, identity, chronology, agency, country, or era.
6. Choose a metric-level winner among SSP, MAS, or tie.

Output ONLY JSON:
\{
  "scores": \{
    "SSP": \{"metric": 1-5, "...": 1-5\},
    "MAS": \{"metric": 1-5, "...": 1-5\}
  \},
  "winners": \{"metric": "SSP|MAS|tie"\},
  "rationale": "short evidence-based rationale",
  "failure\_flags": ["..."]
\}

User payload:
\{
  "independent\_g\_eval\_sample": [sample index],
  "transcript\_reference": [transcript, compacted to 12000 characters],
  "candidate\_texts": \{
    "SSP": [SSP narrative],
    "MAS": [MAS story]
  \},
  "score\_scale": "1=poor, 2=weak, 3=adequate, 4=good, 5=excellent",
  "metrics": [TEXT\_METRIC\_DEFINITIONS]
\}
\end{promptbox}

\subsubsection{Text Metric Definitions}

\begin{promptbox}
\obeylines\obeyspaces
panel\_transition\_clarity:
Signals memory-episode transitions clearly enough for a multi-image sequence rather than collapsing them into one blended summary.

visual\_scene\_extractability:
Can be cleanly broken into concrete visual scenes with identifiable actions, settings, subjects, and material anchors.

macro\_arc\_fidelity:
Preserves the testimony's large-scale arc: major turning points, displacement flow, and before/during/after structure.

identity\_context\_lock:
Makes protagonist identity, era, geography, and social context explicit enough to prevent downstream drift.

testimony\_grounding:
Stays grounded in the testimony's core facts, narrative thrust, and emotional context without unsupported invention.

hard\_anchor\_safety:
Avoids dangerous distortion of dates, places, identity, relationships, and core chronology.

allowable\_fiction\_control:
Creative connective narration remains plausible and does not distort major events, identity, chronology, or agency.
\end{promptbox}

\subsubsection{Image Grouped G-Eval Judge}
This grouped vision prompt was used for the reported image metrics. Each call receives the transcript/source text, condition-specific prompt metadata, and corresponding SSP/MAS images for the same sequence position.

\begin{promptbox}
\obeylines\obeyspaces
You are a grouped vision judge for testimony-specific visual memory artifacts.
Evaluate SSP and MAS images in the same call to reduce scale drift.
Do not reward generic beauty, cinematic polish, or stock refugee imagery by itself.

Image order is fixed by condition labels:
1. SSP
2. MAS

Procedure:
1. Read the transcript/source text and identify concrete visual evidence for this sequence position.
2. Score each image independently on every metric using the same standard.
3. Choose a metric-level winner among SSP, MAS, or tie.
4. Penalize identity drift, era flattening, generic portrait collapse, evidence-free symbolism, and repeated static mood.

Output ONLY JSON:
\{
  "scores": \{
    "SSP": \{"metric": 1-5, "...": 1-5\},
    "MAS": \{"metric": 1-5, "...": 1-5\}
  \},
  "winners": \{"metric": "SSP|MAS|tie"\},
  "rationale": "short evidence-based rationale",
  "failure\_flags": ["..."],
  "visible\_evidence": \{
    "SSP": ["..."],
    "MAS": ["..."]
  \}
\}

User payload:
\{
  "independent\_g\_eval\_sample": [sample index],
  "sequence\_position": "[1..target\_panel\_count]",
  "transcript\_reference": [transcript, compacted to 7000 characters],
  "conditions": \{
    "SSP": \{
      "source\_text": [SSP narrative],
      "image\_prompt": [SSP image prompt],
      "image\_summary": [SSP image summary],
      "image\_filename": [filename]
    \},
    "MAS": \{
      "source\_text": [MAS story],
      "image\_prompt": [MAS image prompt],
      "image\_summary": [MAS image summary],
      "image\_filename": [filename]
    \}
  \},
  "score\_scale": "1=poor, 2=weak, 3=adequate, 4=good, 5=excellent",
  "metrics": [IMAGE\_METRIC\_DEFINITIONS]
\}
\end{promptbox}

\subsubsection{Image Metric Definitions}

\begin{promptbox}
\obeylines\obeyspaces
narrative\_progression\_across\_panels:
Contributes a distinct event, action, or progression rather than repeating static mood or portraiture.

memory\_scene\_shift\_with\_evidence:
Shifts from present interview into remembered time, place, person, or event with textual evidence.

narrative\_action:
Shows action or environmental change rather than only a posed subject or empty scene.

non\_genericity:
Specific to this testimony, avoiding stock refugee imagery, generic sad portraits, or symbolic filler.

testimony\_specific\_scene\_detail:
Contains setting, action, object, social relation, or historical context specific to this interview.

visual\_evidence\_coverage:
Visible elements trace to concrete visual anchors in the testimony/source text.

identity\_continuity:
Race/ethnicity, age logic, gender, and protagonist identity remain consistent unless an intentional time shift is stated.

period\_region\_plausibility:
Era, region, clothing, architecture, and social context are plausible.
\end{promptbox}


\newpage

\section{Case-Level Variable Comparisons}
\label{app:case-variable-comparison}

This appendix provides case-level evidence for how the two pipelines transform the same oral-history testimony into image-generating intermediate representations. For each case, we first show the transcript excerpt from which the outputs were produced, then compare the story representation generated by the Single Summarization Pipeline (SSP) and the Multi-Agent Scene-decomposition pipeline (MAS). We also report the panel-level variables used for image generation. Rather than listing the shared rendering instructions, the tables isolate the case-specific content that differs between conditions: what each system selected as the panel topic and how it specified the visual scene. This makes visible the central contrast analyzed in the paper: SSP tends to preserve the testimony as a compact continuous narrative, whereas MAS makes explicit the remembered scenes, objects, actions, and affective anchors that become available for visual generation.

\subsection{Case 38: Darara Gubo}
\paragraph{Relevant transcript excerpt (45:00--54:00).}
\begin{promptbox}
[45:00 - 45:45] So I share what I read. Probably Uber is gonna experience me, but it's okay, it's fine. If they listen to this interview, they'll be like, we don't want that. That's fine, that's fine, that's fine. So, you... went all in. So you told me that you are running for city council in the city of Clarkston. I did. Yes. So, why... When did you figure out that you wanted to do that... Now politician voice. Yeah. I was, you know, wanted to be a politician.

[45:45 - 46:30] You know, how people, you know, how to fix problems. And one of the most terrifying things is to be in Montana the day of election and to be surrounded with people who are sometimes at liberty to talk about things that I thought are kind of extreme. So it's kind of like there's a feeling of vulnerability as an immigrant.

[46:30 - 47:15] You want to do something about it. You wanna be part of, you wanna be, you know, around the table, you know, and fix it instead of being affected by it, you know what I mean, affected by it. So, coming back to Clarkston and going to Emory and all that, Emory, well, has very, very good programs, you know, reaching out to communities and going to engaging the communities. So I was at one of those events, happened to be in Clarkston, when I met, when I talked to the mayor, he was there too, and.

[47:15 - 48:00] I expressed my desire to be part of the city, and he was very welcoming. He was so eager to have me there. And Mayor Tate, who is now running for Senate, is really great, great guy. He wants to elevate us, the immigrants. So when we started talking, I was not even a citizen yet. I became a citizen on July 10th. Today is July what... 24th... 24th, so two weeks. Congratulations! Thank you so much. Yeah, so I started to think about it, and then I had, you know, when I was talking to him, I already had applied for my citizenship.

[48:00 - 48:45] to support going through the channels, whatever. So I started to really think about it, and I also went to a meeting where the city council was holding a public hearing. It was about some building, and there was so much passionate debate about it. And I thought the majority of people were on the wrong side of the debate, from my perspective. So I stood my ground, and all those things showed me that I have to really, really be part of the government to contribute and, you know,

[48:45 - 49:30] So I started to talk to people and community members and all that. I remember talking to this guy who really is well-connected in the community. He's like, don't start out for city council, run for commissioner, be commissioner, you know, he's an immigrant. He said, we want an immigrant to be the county commissioner. It's a big office. But I feel like, I don't wanna be far away, I wanna be here. I wanna be like with you guys. Close to the people. Yeah, you know, maybe one day, but who cares... I mean, my point is not to be up there. I wanna be just down here with you guys. And he's like, but you know, he has a very strong point to make because he's like, we want...

[49:30 - 50:15] Yeah, we need someone to advocate there. Maybe one day, or maybe somebody else. But for me, you know, you just have to have that, you know, you have to have that personal calling. You know what I mean... As we say in our first tradition, God has to call you to do something. And I feel like God is calling me to do City Council and work with the city, work with, you know, my community, my neighbors. You know what I mean... Yeah. And my goal is to know everyone in my city. It is a small town, so thankfully it's achievable. Definitely. And to visit them and to get to know what their deal is, what the struggles are, if it's a work, figuring out.

[50:15 - 51:00] We'll find that connecting them with organizations or all that. If it's about medical care, because thankfully the collection has a lot of nonprofits working in the community, so connecting them with organizations or if it's a new American pathway, if somebody is trying to be a citizen, connecting them with a new American pathway. If they want to register for a vote, you know, for me, when I registered because of this organization, when I registered for voting, I came here and said, help me to fill out the form, the new American pathway. So I do believe if we care enough for the people, and if we spend enough time with people, we can...

[51:00 - 51:45] meet their needs. Because there is a need here, and there is a resource here. So the role of a politician, I should be, I think, to figure out how to make both meet. You know what I mean... Not always through government channel, not always through making funds available. It could be to guide them to go to organizations. It could be urging them to go to different religious community organizations. So it could be Christians, it could be Muslims, it could be Hindus, it could be whatever organizations out there connecting the most people so that they can meet the needs of the people. So do you feel like your faith will help you in the situation that

[51:45 - 52:30] You are elected to the city council... Definitely. That's why I'm running. I'm running because of my faith. My faith says, if you see Lord, Lord, or if you see Jesus, Jesus, and if you don't do the will of God, he says, I don't know you. It's in the Bible. I don't know you. Saying, talk is cheap. You know what I mean... Talk is cheap. So go help people. Go help those who are in prison. Visit them and go to those who are naked and close them, who are homeless, you know, which is, there was a point in my life where I found myself.

[52:30 - 53:15] living in a car actually, so I know what that means and how that feels, you know. So my election is all about fighting for the poor, for the middle class, for people who are struggling, and making sure that they get the bite out of, you know, the American dream. And by the American dream, I mean, you know, decent lives, you know, decent life, affordable living situations. So, yeah. So what is the American dream for you... Um, yeah, just, you know, people having a place to sleep, clothes, to have clothes, and to have medical care, you know, you know, those big things.

[53:15 - 54:00] I would say that for me that's American dream. If it's about piling billions of dollars, millions of dollars, making a bigger house at the expense of the poor, I'll call that American greed. Wow, that's powerful. Yeah, so we gotta figure out how to help each other. Not necessarily through forcing people to help. It's not working. It doesn't work. So that's why I tend to work with community organizations and all that to help meet the needs.
\end{promptbox}

\newpage

\paragraph{Story.}
\begin{longtable}{>{\raggedright\arraybackslash}p{0.47\linewidth}>{\raggedright\arraybackslash}p{0.47\linewidth}}
\toprule
SSP & MAS \\
\midrule
{\raggedright\footnotesize\ttfamily Darara Gubo recounts her transition from Ethiopia to the United States and her evolving engagement in local civic life in Miami-area communities. She describes early culture shock and vulnerability as an immigrant, observing polarized public speech during a Montana election and later participating in community outreach events in Clarkston and at Emory. While not yet a citizen during initial outreach, she applied for and received citizenship on July 10, which intensified her desire to influence local decisions. At a contentious city council hearing she felt compelled to speak and, encouraged by community members and a mayor who welcomed immigrant participation, she resolved to seek local office. Darara emphasizes a preference for staying close to neighbors rather than pursuing distant power; her stated calling to serve city council comes directly from her Christian faith and past experiences of poverty and instability.} & {\raggedright\footnotesize\ttfamily Scene 1: In the present interview room the speaker recalls fear at being in Montana on election day, surrounded by people speaking extreme things; must\_preserve: speaker, election-day fear, surrounding hostility; may\_imagine: chair, recorder, window light and restrained gestures. Scene 2: Remembered conversation with Mayor Tate, who welcomed the speaker and spoke of elevating immigrants while the speaker was not yet a citizen; must\_preserve: welcome, Mayor Tate, speaker not a citizen; may\_imagine: warm office light and a handshake. Scene 3: A community connector urges the speaker to consider running for commissioner rather than council; must\_preserve: adviser, recommendation, adviser is an immigrant; may\_imagine: neighborhood meeting, flyers. Scene 4: The speaker thinks through citizenship channels and linking newcomers to nonprofits  Time for the Collection and medical referrals are mentioned; must\_preserve: connecting immigrants to organizations and services; may\_imagine: visiting a nonprofit desk and paperwork. Scene 5: The speaker says faith motivates the campaign, quoting biblical admonition about doing Gods will; must\_preserve: faith as motive and scripture reference;...} \\
\bottomrule
\end{longtable}

\newpage

\paragraph{Image prompt variables.}
\begin{longtable}{>{\raggedright\arraybackslash}p{0.08\linewidth}>{\raggedright\arraybackslash}p{0.43\linewidth}>{\raggedright\arraybackslash}p{0.43\linewidth}}
\toprule
Panel & SSP variables & MAS variables \\
\midrule
1 & {\raggedright\footnotesize\ttfamily Summary: Arrival and first impressions in Miami  an immigrant woman steps off a bus or train into a humid, sunlit city and feels disoriented.\par \par Variable visual brief: Wide-angle scene of a lone immigrant woman of Ethiopian origin stepping off a bus into a bright, humid urban Miami streetscape. Palm trees, low-rise buildings, and warm golden light create strong contrasts. She holds a small suitcase and a tote bag with papers peeking out, wearing modest professional clothing and an expression of thoughtful uncertainty. Background figures bustle, some glancing at her, conveying a mix of welcome and anonymity. Cinematic, high-detail, naturalistic lighting, shallow depth of field to emphasize her face and posture.} & {\raggedright\footnotesize\ttfamily Summary: Present-day interview room showing the speaker remembering fear on election day in Montana, surrounded by hostile conversation.\par \par Variable visual brief: Interior interview room, midshot of a thoughtful woman (no identifying labels) sitting in a simple chair, hands restrained on her lap, expression tense and watchful. A small voice recorder and a notepad on a table in front of her. Through a window, muted blue Montana light. Background blurred figures seen through glass or reflected speech bubbles suggesting hostile words and extreme slogans (abstract, not legible). Atmosphere: quiet tension, cinematic documentary style, natural window light, soft shallow depth of field.} \\
\midrule
2 & {\raggedright\footnotesize\ttfamily Summary: Spiritual solitude described as a 'desert'  the woman reflects on cultural and spiritual isolation in a new country.\par \par Variable visual brief: Intimate portrait of the same woman seated alone on a bench near a small, sparse urban park at dusk, symbolic 'spiritual desert' mood. Cool blue and purple tones, long shadows, and a single streetlamp create a quiet, contemplative atmosphere. She clasps a small Bible or notebook to her chest, eyes closed or looking into the distance, conveying introspection and spiritual searching. Subtle textures, emotive lighting, painterly realism.} & {\raggedright\footnotesize\ttfamily Summary: Recalled meeting with Mayor Tate where she was welcomed despite not yet being a citizen.\par \par Variable visual brief: Warm office scene, mayoral office ambience, midshot of a handshake between the woman and a man representing 'Mayor Tate' (no nameplates). The woman looks hopeful but slightly reserved; the mayor smiles and gestures in welcome. Sunlit warm tones, framed certificates and a city seal suggested on the wall but not named. Emphasize welcome, inclusion, and that the woman is not yet a citizen through subtle visual cues like a thin stack of immigration forms on the desk. Documentary realism, soft cinematic lighting.} \\
\midrule
3 & {\raggedright\footnotesize\ttfamily Summary: Witnessing polarized public speech during an election  watching heated political rhetoric on TV while processing its impact.\par \par Variable visual brief: Interior scene of a modest living room where the woman watches a television broadcast of a heated political debate or campaign speech. TV shows two contrasting speakers animated and gesturing; the room is dimly lit by the screen's glow. The woman's face reflects concern and surprise; newspapers and community flyers are scattered on a coffee table. Insert visual cues of civic engagement: voter registration forms, a map, sticky notes. Photo-realistic, moody lighting, close framing on face and screen.} & {\raggedright\footnotesize\ttfamily Summary: A community connector, an immigrant adviser, urging the speaker to consider running for commissioner instead of council.\par \par Variable visual brief: Neighborhood meeting scene in a community center: a mid-aged immigrant adviser leaning forward, animatedly encouraging the woman, pointing toward a simple campaign sketch on a table that reads 'Commissioner' (generic). The woman listens thoughtfully. Around them, diverse community members, flyers and a bulletin board with multilingual notices. Warm, collegial atmosphere, candid documentary photography style, natural indoor lighting.} \\
\midrule
4 & {\raggedright\footnotesize\ttfamily Summary: Participation in community outreach events  engaging with neighbors at local gatherings and campus activities.\par \par Variable visual brief: Lively community outreach scene at a neighborhood event: the woman converses warmly with diverse neighbors at a table with informational pamphlets, snacks, and a poster that reads 'Community Outreach' (generic). Another vignette shows her speaking with students on a university campus lawn, folding chairs and booths in the background. Bright daylight, inclusive crowd, candid expressions of dialogue and connection. Vibrant colors, documentary-style realism, medium shot composition.} & {\raggedright\footnotesize\ttfamily Summary: The speaker thinking through citizenship pathways and linking newcomers to nonprofits and services like "Time for the Collection" and medical referrals.\par \par Variable visual brief: Close-up scene of the woman at a nonprofit desk sorting paperwork: citizenship forms, a pamphlet titled 'Time for the Collection' (generic), and a list of medical referral contacts. A volunteer at the desk gestures helpfully. Visual cues of connection: a small map with arrows to local organizations, business cards, clipboards. Soft hopeful light, organized clutter, realistic documentary aesthetic.} \\
\midrule
5 & {\raggedright\footnotesize\ttfamily Summary: A pivotal moment at a tense city council hearing  she stands at a microphone to speak, supported by community members and a welcoming mayor.\par \par Variable visual brief: Dramatic interior of a city council chamber during a contentious hearing: the woman stands at a microphone in the public comment area, mid-speech, composed and determined. Behind her, a mixed group of neighbors offers encouraging looks and a mayor or councilmember nods supportively from the dais. Tension in the chamber is visible: some attendees fold arms, others take notes. Warm, directional lighting highlights the speaker; formal wood-paneled architecture, flags, and a hearing agenda on a wall. Realistic, cinematic documentary style.} & {\raggedright\footnotesize\ttfamily Summary: The speaker explaining that her Christian faith motivates her public service, referencing scripture and moral duty.\par \par Variable visual brief: Intimate portrait of the woman speaking earnestly, hands slightly raised in a meaningful gesture, a small open Bible or scripture passage visible on a table beside her (no specific translation or church branding). Warm, reverent lighting like late afternoon sun. Background minimal to focus on expression of conviction. Tone: sincere, calm, inspirational documentary style.} \\
\midrule
6 & {\raggedright\footnotesize\ttfamily Summary: Commitment to local service rooted in faith and past struggles  she resolves to run for city council focused on neighbor-centered leadership.\par \par Variable visual brief: Uplifting, hopeful portrait of the woman in a neighborhood setting standing on a front stoop or small community garden, looking confidently toward the street where neighbors pass by. She holds a simple campaign-style notecard or flyer reading 'For Our Neighborhood' (generic), and wears a small cross necklace as a subtle sign of faith. Children and elders in the background, warm morning light, soft-focus background to emphasize her resolve. Inspirational, naturalistic color palette, high-resolution detail.} & {\raggedright\footnotesize\ttfamily Summary: A hopeful campaign moment showing community outreach and the speaker linking immigrants to services while preparing to serve publicly.\par \par Variable visual brief: Outdoor community outreach scene: the woman standing at a mobile information table handing flyers to diverse newcomers, volunteers assisting with sign-up sheets and medical referral forms. A banner reading 'Community Resources' (generic) draped behind. Faces show determination and hope. Bright daylight, vibrant but realistic colors, wide-angle documentary shot capturing community energy and purpose.} \\
\midrule
\bottomrule
\end{longtable}

\newpage

\subsection{Case 80: Jorge Quintas}
\paragraph{Relevant transcript excerpt (01:15:00--01:35:15).}
\begin{promptbox}
[01:15:00 - 01:15:45] Cross the border and go into the United States. So she got in function of that. She talked to my dad and managed to get my dad a reunification visa, like family reunification. So after three months over there, my dad, my sister, and I, finally all of us were like in Mexico. And then we started planning the trip. And we had a cousin that was living in Venezuela at the time, and he also wanted to come to the United States. So our plan was the following. We go, all of us meet, like, he takes a plane from Venezuela.

[01:15:45 - 01:16:30] to Colombia, from Colombia to Reynosa, which is pretty much a border town in Mexico. Like on the other side of the border, you have McAllen, Texas. So we were planning to do that and then from Cancun, we would just take the plane straight out to um McAllen, not McAllen, but like the Mexican part of McAllen, which is Reynosa. And we would just meet up there and then all of us would like get on a taxi and like cross the border at the same time. Um The border is 15 minutes away from the airport, and my mom had known people who had done this in the past, so. So we were pretty confident in the plan.

[01:16:30 - 01:17:15] However, when my cousin went to take a trip from Venezuela to Colombia, he was held up because of paperwork, and Colombia like required a visa, so like he couldn't make it. And then we were like desperate because like money was running out in Mexico, and then we had like family in the United States that was going to be able to like receive us pretty much and take care of us. So we just had to go on ahead and continue without my cousin. He later came. So we traveled, we took the airplane, we traveled from Cancun to Reynosa. But the thing is,

[01:17:15 - 01:18:00] South Americans and Mexicans really hate Cuban people. They hate them because the law that was implemented before, like the wet food, dry food. If you're a Cuban and you get onto American soil, you become a resident instantly. And then you have all the rights. You get given food stamps, you get money. If you don't have a house, you can tell them and they will find you some house somewhere. It may be in the middle of Oklahoma, but they're gonna find you a house. And the thing is, the rest of South America wants to come to the United States and they have to struggle a lot to travel all the way up to the United States.

[01:18:00 - 01:18:45] States, and then once they're in the border, they just, like, get rejected. So like, we're really hated amongst the South Americans for that, and when we were living in Mexico, like, sometimes I would show. And the stories like other people told us, like, working as a Cuban in Mexico, like, the employers just, like, treat you very badly and the people treat you badly. I never got to experience it, but uh But it's true, I did see it. And yeah, when we got to the airport in Reynosa, we got out of the airplane, and the first thing that we saw was like, going through immigration, one of the immigration officers.

[01:18:45 - 01:19:30] We hand him our Cuban passports, and then he's like, hey, you should come out into this little room, so like, for like further checking pretty much. And then like once we got inside like in the little room, um he like started talking to us and he was like, hey, you're Cuban, you are here because you're trying to cross the border. Pay me money or I'll deport you. And uh We did not have a lot of money because we had already been living in Mexico for like a little bit and and yeah, we just did not have a lot of money. However, we were smart enough to like hide some of the money and after like talking to this man, like, hey, we're not trying to do this. We made up a story that like someone.

[01:19:30 - 01:20:15] was gonna come from the United States to see us here and buy us a car. So that was like our official story. He did not care, and he just wanted money. And after I talking to him for a while, it was okay, whatever. You know what... Take the money. This is all the money that we have. And we like gave him like a thousand Mexican dollars, which is like nothing pretty much. And then he looked at the money, he was like, is this all you have... And then he gave us half of it back. He was like, no, man, take this. You're gonna need it. he felt pity for us for the little money that we have. The other money was hidden, but like, it was not much more. So after we paid him, we get out of the airport. And just like walking out of the airport, you get a sense that it's a different city.

[01:20:15 - 01:21:00] And you get a sense that it's a border town because like right outside of the gates of the airport, you have like two officers like with face masks, like holding like AR-15s, just like on the entrance of the airport, so like make sure like, I don't know, nothing happens. So since this is a border town, where there's like a lot of like trafficking going on, and human trafficking especially, so my mom told us, like, hey, let me do the talking and then don't talk, because it's like by just talking your Cuban Spanish, like people can tell that you're Cuban. So we got in, we get

[01:21:00 - 01:21:45] In a taxi. Also, I think something else that gave us away, we were just like holding a lot of luggage, like a whole lot of luggage for each person. So like, if you're just gonna come here to visit for a little bit, I don't think that you want to get that much luggage with you. So we got a taxi, we like put all the luggage in the trunk. It was so much luggage that the trunk didn't even close. And then we all get inside the same taxi. It was a little crammed, but like we all fit. And then the guy was like, Hey, where are you going... And we just I'm like, yeah, just drive. We'll tell you later. We're trying to find the name of the hotel that we're going. We were faking. And then he just starts driving. And then like my mom's Mexican accent was not that.

[01:21:45 - 01:22:30] So like a little bit of Cuban slipped through, and while we're driving down this long road away from the airport, the taxi driver gets a phone call, and he picks it up. He says, yes. He looks at us and says, yes. And he's like, OK, and then he hangs up the phone. And he stops the taxi on the side of the road. And he's like, hey, the cartel just called me. Up ahead, we're going to be stopped by the cartel, and you're going to have to pay \$10,000 for each one of you. If not, they're going to take you. Do not get me involved. I do not wish to get involved. I will be killed. I have a family. I don't care for you guys.

[01:22:30 - 01:23:15] And then like we just like freaked out because that's not something that you hear very often. The cartel is gonna stop you up ahead and they're gonna ask for money, and if not, they're gonna take you. So like we tried like, we started trying to negotiate with this guy and like what do you mean like \$10,000... Like we did not have that money. This man is like not taking it. He's like, I don't care. You talk to the guys when you get there. Do not get me involved. I don't wanna be killed. That's something that he said repeatedly, like, I don't wanna be killed. So we're like, OK, just like go ahead, I guess. Because we're in the middle of nowhere. It was just like one long strip of road and then just like nothingness all around. So we get to the

[01:23:15 - 01:24:00] we get to the end of this like long road, and there's like a T intersection, and to the right of the road, there was a like a small mall with like a 7-Eleven, a gas station, a bank, and some like other retail stores. And we, we see this and like my mom goes like, hey, can you like stop here, like inside this little mall, so like we can go to the bank and get out some money because we don't have this money with us. And he bought it, so he went, he goes, he parked inside the, inside the mall, and as soon as he parked, we just, all of us like commando, we just get out of the car, go to the back, which was not closed because we had so much luggage, open up the trunk, and then we

[01:24:00 - 01:24:45] We take out of her luggage and then we ran inside the 7-Eleven. And at this point, you're at a crossroads, like you don't know what to do, like, what just happened... Like someone just said that we're gonna get kidnapped if we don't pay the cartel money. And then I guess we're trying to process what's going on. You know how a 7-Eleven is, like the walls are like made out of like crystal. Yeah, like windows, like windows all around the 7-Eleven. You go and like, you can see everything that's going on outside. And then like I was trying to figure out what's going on, a van comes and pulls out and like just parks like outside.

[01:24:45 - 01:25:30] to the 7-Eleven. And like two like mean Mexican-looking guys like just park outside, then they get out and they just like get in front of the car and they just like cross their hands and they just like start talking to each other and looking at us. And then after like another one shows up and at some point we have like five or six different vans outside like doing the same thing. And then we're like, oh, shoot, what are we gonna do now... So this is a very scrawny, small Mexican guy walks in and he walks in with all the confidence in the world. He just like, by the way, before like this guy walked in.

[01:25:30 - 01:26:15] Like we were talking to like the people in the 7-Eleven telling us like what was going on. The people freaked out. They told us don't speak with us, you're all gonna get killed, and then they just left the 7-Eleven to ourselves. They went and they hid in the back. And we were like, don't talk to us, you're gonna get killed, we have families. So we're like, cool, we don't even know where we are. Yeah, it's fine, this is fine. And then like this little like Mexican guy just comes in and he just like calls my dad over. My dad is like 230 pounds at the time, 5'8". And then you have this little like Mexican guy like.

[01:26:15 - 01:27:00] I came over and like talking to him and be like, hey, we're gonna take care of you, like, don't make this any harder than it has to. Um we're very good people, like, you know, like all these different bands that are outside are different groups that are trying to get you to go with them. However, you should come with us, like, we're gonna treat you well. Like, as soon as your family pays the money, like, we're gonna let you, we're gonna, we're gonna help you cross the border, and like, my dad is like, we do not want to cross the border. What do you mean you're gonna treat us well... Get out of here. Like, what do you mean things are gonna get bad... Like, who's gonna, like, who's gonna drag us out of here... You're gonna drag us out... And then like the guy just like looks at him and he's like, no, I'm not gonna drag you out of here.

[01:27:00 - 01:27:45] Who's gonna drag you out here is gonna come later, don't worry. And then like this little man threatening my dad, and yeah, like eventually he left and my mom all of a sudden, she was trying to call the Mexican police. She managed to get like out of one of the workers, like the name of the place, the place was called like Plaza Aeropuerto, which is pretty much like Airport Mall. Okay, makes sense. We tell the Mexican police that. And then in the distance, three army trucks pulled out, like we assumed that they were here to get us. And the trucks had, one of the trucks had like a mounted.

[01:27:45 - 01:28:30] a mounted gun on top, and everybody was like covered, and everybody was like carrying guns. And they saw us inside the 7-Eleven. They saw all the different bands outside, and then they turned around and they left. And my mom was still on the phone with the police, and she was like, hey, like, what happened... Like, these people just left. And the police were like, oh, these people are bought, my bad. So pretty much these people have been bribed, like, by the cartels, like, to look the other way whenever something like this was happening. So my mom, at that point, she knew some people who like, had crossed the border before and who had relations with people who crossed the border, and coyotes, and she called them, and they told her to.

[01:28:30 - 01:29:15] We call them the Mexican DEA pretty much, because that's the only thing that like, that those are the only people who are not bought by the cartels once that they're there, and who just like fight the cartel all the time. The only thing that the cartel are scared of. So we did that, we tell them who we are, like where we are. And after a while, like, it felt like an eternity pretty much. The Mexican DEA pulls up. Same setup. Three different trucks, everybody face covered, um, glasses, uh hats, uh carrying like long guns, and on three cars with like one of them had a mounted gun. Two of them actually had mounted guns on top.

[01:29:15 - 01:30:00] And once that they pulled up into the plaza, the Mexican, the Mexican, the people who were outside like trying to kidnap us and like um once they saw who they were, they skedaddle. Like all of them got inside their cars and they just like started burning rubber and they just left the place. Um and these people come inside the 7-Eleven, they get us and they like start talking to us, like, hey, what happened... We told them like, we were just like trying to get to a hotel and then like someone in the taxi, like called the taxi driver and the taxi driver like told us this. And he's like, okay, no problem. Like, what do you guys wanna do now... He's like, could you take us to like a hotel so that.

[01:30:00 - 01:30:45] we can like process what's going on and like, he's like, sure, no problem. They load us inside, me, my sister, and I, they put us inside one of the trucks. And then my dad later told me that like, because they didn't have space in the same truck, they put him in another one. But to get him inside the truck, they had to move like boxes of ammunition and grenades to the back of the truck because it was filled to the brim with like boxes of ammunition. And then we're driving towards the hotel, and then when we're driving, we see a taxi. And like, the guy that was in command, like he asked us, is that the taxi... We're like, we don't know, like all the taxis look the same, right... And then he's like, no problem.

[01:30:45 - 01:31:30] He gave like a command over the radio, and then the three trucks, one of them parked in front of the taxi, the other one on the side, and the other one in the back. So they pretty much enclosed him, stopped him in the middle of the freeway, and everyone like got out. Like, it was guns blazing. They went up to the taxi driver and they grabbed him by the neck. Like, he had his window down. They just grabbed him by the neck, and they pulled him outside of the car by the neck, through the window. And then they get him up to like my mom. No, they get him to my dad, and he's like, is this the guy that was like driving the taxi... My dad was like, I don't know, I was sitting in the back, go ask my wife. And they just like racked him like a ragdoll. And they just like bring him over to where my mom was. And he was like, is this the guy... She was like, no. And then they leave the guy there.

[01:31:30 - 01:32:15] Oh, my bad. How you doing... You okay... Okay, carry on. I said carry on. Um, and that happened. And yeah, these people meant business. They, um, when we were, when they were trying to like get the guy out of the car, one, a small, small Mexican lady who was there, who was the second in command, we later found out. Um, she's like, hey, get low to the ground because like half of the vehicle, the bottom half is like bulletproof. So just in case like someone comes driving and like does a drive-by, so that you guys will be safe. And we're like, OK, to the ground, OK, to the ground. So, um, eventually we get to the hotel and the hotel.

[01:32:15 - 01:33:00] The hotel they took us to was the same hotel that the entire like Mexican DEA was staying in for that town. So now we're in the hotel, we book a room, and we just like get inside and we start to process what's going on. We're like, okay, we're trying to cross the border illegally. We almost got kidnapped. Now we're in a hotel full of policemen pretty much who could like deport us if anything like goes wrong. Yeah, what are we gonna do now... And we had brought with us like a bottle of like wine. Like it was like a humongous bottle of wine. It was like a small baby. That's how big it was. And we also had brought a Cuban tobacco.

[01:33:00 - 01:33:45] So we had the phone numbers of one of the guys who, like, had rescued us, and we called, we hit him up, and he's like, hey, what are you guys doing tonight... Are you free... Like, could we talk... And he's like, sure, no problem. Like, right after we come back from patrol, we can just, like, go out and, like, have a drink. I was like, a drink... OK, OK, that sounds good. So they came back around, like, 9 p.m., and they took us, like, across the street from the hotel to a bar, a karaoke bar, nonetheless. And they closed down the entire bar so that we and them could be there. Like, they just closed down business for the day. And then these people just, like, churned out.

[01:33:45 - 01:34:30] 180 degrees. They were no longer like these hardcore military guys like carrying big guns. These people were just like drinking and dancing and drinking tequila and like doing karaoke. And like, it was very fun. And after like celebrating with them for a while, my dad goes like, hey, we were here to like, so that my sister could come over and see us and like buy us a car. However, after telling her what happened, she's a little scared. And we were thinking like if maybe we could like go over there, like, we didn't know how these people were gonna react to this, or we don't know if they were gonna deport us. And this guy like looks at my dad, takes a shot of tequila, and he's like, no problem. Tomorrow, I got you.

[01:34:30 - 01:35:15] And it was like, okay. And then like, we just continued like celebrating, I guess. And the next morning we woke up around 11. We had breakfast and then they came, they came back from patrolling because even though like they just got drunk, very drunk, the night before and they still had to get up for patrol at 6 a.m. and they did that. No problem. We had breakfast and then they, they made us call over a taxi. The taxi came and then they talked to the taxi driver, like the one who was in command went over and he's like, Hey, do you see those people over there... You're gonna take them to the border. Now, these three cars over here are gonna be scorting you. Do not get away from them.
\end{promptbox}

\newpage

\paragraph{Story.}
\begin{longtable}{>{\raggedright\arraybackslash}p{0.47\linewidth}>{\raggedright\arraybackslash}p{0.47\linewidth}}
\toprule
SSP & MAS \\
\midrule
{\raggedright\footnotesize\ttfamily Jorge Quintas recounts his familys escape from Cuba through Mexico toward the United States. Traveling by air from Cancun to Reynosa with heavy luggage and a delayed cousin, they immediately sensed hostility toward Cubans in Mexico and encountered corruption at the airport: an immigration officer demanded money and returned most of a small payment in pity while unseen cash remained hidden. On the road a taxi driver warned of a cartel ambush demanding ransom. His mother improvised, steering them into a mall where vans and men surrounded them. A small Mexican intermediary tried to recruit them; threats escalated until Jorges mother contacted authorities. After an initial army presence withdrew, the family called the Mexican DEA, who arrived in armed convoys and dispersed the wouldbe kidnappers. The DEA escorted Jorge, his sister and later his father to a hotel where they processed the ordeal. That night the agents relaxed, socialized with the family in a closed karaoke bar, and one promised to arrange safe passage for the sister. Throughout Jorge describes fear, bargaining, and relief, noting how danger, bribery, and uneven protection shaped their migration.} & {\raggedright\footnotesize\ttfamily Scene 1: In the present interview room, the speaker sits restrained, voice low, and describes how a cousin travelling from Venezuela to Colombia was held up by visa paperwork while the family ran out of money in Mexicomust\_preserve: cousin's trip and the paperwork delay, growing desperation; may\_imagine: soft window light, the speaker's hands clasped. Scene 2: Memory shifts into a border-town transit lane: the recalled subject walks past an airport gate where two masked officers stand with rifles; the narrator names hostility toward Cubansmust\_preserve: armed officers outside airport and sense of border hostility; may\_imagine: dust, distant engines, tense crowd. Scene 3: The remembered family approaches a T-intersection and a small mall with a 7-Eleven and gas station; the mother calls for help and they hurry toward the storesmust\_preserve: mall/7-Eleven landmark and mother's prompting; may\_imagine: screeching tires, hurried footsteps. Scene 4: Inside a small processing room the family hands over Cuban passports; an agent says to step into a side room for further checks while offering calming wordsmust\_preserve: handing Cuban passports, being called into a small room for...} \\
\bottomrule
\end{longtable}

\newpage

\paragraph{Image prompt variables.}
\begin{longtable}{>{\raggedright\arraybackslash}p{0.08\linewidth}>{\raggedright\arraybackslash}p{0.43\linewidth}>{\raggedright\arraybackslash}p{0.43\linewidth}}
\toprule
Panel & SSP variables & MAS variables \\
\midrule
1 & {\raggedright\footnotesize\ttfamily Summary: Arrival in a Mexican border airport: family with heavy luggage encountering corruption and hostility toward Cuban travelers.\par \par Variable visual brief: Interior of a busy regional airport at dusk, a family of four with heavy suitcases looking anxious at a passport control desk. Focus on tension: a stern immigration officer in uniform pockets money discreetly while returning a small wad of bills with a pitying expression. Other passengers glance with suspicion; signs in Spanish in the background, fluorescent lighting, weary faces, subtle film grain, cinematic close-up of hands exchanging cash, muted color palette, emotional realism, photo-realistic style.} & {\raggedright\footnotesize\ttfamily Summary: Present-day interview room: restrained speaker, low voice, anxiety about a cousin delayed by visa paperwork and family running out of money in Mexico.\par \par Variable visual brief: Interior interview room with a single table and two chairs, a middle-aged man seated and slightly restrained at the wrists, voice-muted expression, hands clasped in his lap. Soft window light falls across his face, creating a quiet, intimate mood. On the table, a folded photograph and a worn travel document hint at a cousin's trip delayed by visa paperwork and growing family desperation. Keep focus on the subject's tense posture and worried eyes, muted color palette, shallow depth of field.} \\
\midrule
2 & {\raggedright\footnotesize\ttfamily Summary: Taxi driver's warning about cartel ambush while driving through a tense highway toward Reynosa.\par \par Variable visual brief: Nighttime highway scene through a car windshield: a worried mother and two children in the back seat, a tense driver in the front warning them with a gesture. Roadside shacks and distant headlights streaking past, a billboard in Spanish, heavy luggage visible in the trunk. Foreboding atmosphere, dramatic low-key lighting, rain-slicked road reflections, close framing capturing fear and urgency, cinematic realism.} & {\raggedright\footnotesize\ttfamily Summary: Memory: border-town airport gate with two masked officers armed and a tense crowd, explicit hostility toward Cubans.\par \par Variable visual brief: Outdoor airport transit lane in a dusty border town at midday, travelers milling past a gate; two masked officers stand guard in front of an entrance, uniforms and rifles visible, posture alert and imposing. Crowded scene with a tense atmosphere, whispered hostility expressed in body language and distant shouted words. Include dust motes, distant airplane engines, and anxious travelers looking over their shoulders. Emphasize sense of border hostility toward Cubans without showing identifying text on people.} \\
\midrule
3 & {\raggedright\footnotesize\ttfamily Summary: Improvised escape into a shopping mall as vans and armed men surround the family.\par \par Variable visual brief: Exterior of a mid-sized shopping mall at night, family hurriedly entering through automatic doors while several unmarked vans pull up and men in plain clothes exit, forming a semicircle. Tension and chaos: frightened faces, menacing body language, mall lights contrasted with dark vans, bystanders frozen, dramatic wide-angle composition, high contrast, intense cinematic mood, photorealistic detail.} & {\raggedright\footnotesize\ttfamily Summary: Family hurrying toward a T-intersection and small mall with a 7-Eleven and gas station as the mother calls for help.\par \par Variable visual brief: Street-level view of a T-intersection leading to a small mall: storefronts visible with a recognizable 7-Eleven-style convenience store and adjacent gas station canopy. A family hurries across the pavement, mother calling out and gesturing toward the stores, hurried footsteps and blurring motion. Add environmental details: screeching tire sounds suggested by motion lines, scattered trash, late-afternoon light. Convey urgency and the mother's protective command.} \\
\midrule
4 & {\raggedright\footnotesize\ttfamily Summary: Arrival of Mexican federal agents (DEA equivalent) in armored convoys dispersing the would-be kidnappers.\par \par Variable visual brief: Armored convoy of federal agents in tactical gear arriving at the mall parking lot at night, headlights and flashing emergency lights cutting through smoke or dust. Agents moving in formation, detaining or dispersing the plainclothes men; visible relief on the family's faces as they are guided to safety. Dynamic action composition, gritty realism, strong directional lighting, documentary photo style, cinematic urgency.} & {\raggedright\footnotesize\ttfamily Summary: Inside a small processing room where the family hands over Cuban passports and an agent invites them into a side room for further checks while offering calming words.\par \par Variable visual brief: Cramped processing room with a plain desk and fluorescent lighting; family members seated, handing over visibly stamped passports labeled generically as travel documents (no sensitive personal data visible). An official gestures toward a small side room and speaks in a calming manner, palms open in reassurance. Close-up on the exchange of passports, nervous fingers, and the doorway leading to the side room. Tone is tense but the agent's body language is meant to soothe.} \\
\midrule
5 & {\raggedright\footnotesize\ttfamily Summary: Aftermath at a hotel: agents and family in a closed karaoke bar, agents socializing and promising to help arrange safe passage for the sister.\par \par Variable visual brief: Interior of a small private karaoke room in a modest hotel late at night: the family seated on a sofa opposite a few relaxed agents in uniforms, a microphone on the table, soft colored lights, weary smiles and guarded camaraderie. Intimate, warm but uneasy mood, details like hotel key cards and travel bags visible, shallow depth of field, photo-realistic, human-focused portraiture capturing complex emotions of relief and lingering fear.} & {\raggedright\footnotesize\ttfamily Summary: Side-room check: heightened tension as family waits, documents examined, and the cousin's visa delay causes growing desperation.\par \par Variable visual brief: Narrow side room with a single chair and a small table where an agent examines documents under a lamp; family clustered in the doorway, faces showing fatigue and worry. On a bench nearby, an empty wallet and a small pile of coins suggest the family running out of money in Mexico. Visual cues include a calendar on the wall with crossed-off days and a folded airline ticket marked 'delayed' to imply the cousin's trip stalled by visa paperwork. Low-key lighting, close emotional framing.} \\
\midrule
6 & {\raggedright\footnotesize\ttfamily Summary: New life in Miami: the young narrator as a college student, navigating American education while thinking about the future.\par \par Variable visual brief: Daytime campus scene in Miami near palm trees: a young man carrying a backpack and textbooks, standing on a college quad looking hopeful, city skyline and palm-lined streets in the background. Casual modern clothing, sunlight, warm color grading, subtle hints of multicultural neighborhood, documents tucked away (no visible IDs), optimistic expression, candid documentary style, high realism.} & {\raggedright\footnotesize\ttfamily Summary: Later: Miami college-student life in Hialeah  narrator studying, navigating American education and planning for the future.\par \par Variable visual brief: Bright college classroom or study area in Miami with palm trees visible through the window, a young man taking notes at a desk, textbooks and a laptop open, attentive and hopeful expression. Subtle cues to Hialeah: bilingual flyers on a corkboard (English and Spanish), casual urban clothing. Mood is forward-looking and determined, natural daylight, medium shot showing the narrator integrating into academic life and planning for the future.} \\
\midrule
\bottomrule
\end{longtable}


\end{document}